\documentclass[aps,pre,twocolumn,amsmath,amssymb,nofootinbib,superscriptaddress,floatfix,eqsecnum]{revtex4-1}

\usepackage{amsmath,amsfonts,amssymb}
\usepackage{amssymb}
\usepackage{amsthm}
\usepackage[dvipdf]{color}
\usepackage{graphicx}
\usepackage{dcolumn}
\usepackage{bm}

\begin{document}
\title{Switchable phonon diodes using nonlinear topological Maxwell lattices}

  \author{Di Zhou}
\email[Corresponding author: ]{dizhou@umich.edu}
 \affiliation{
 Department of Physics,
  University of Michigan, Ann Arbor, 
 MI 48109-1040, USA
 }

  \author{Jihong Ma}
 \affiliation{
Department of Civil, Environmental, and Geo- Engineering, University of Minnesota, Minneapolis, MN 55455, USA
 }

  \author{Kai Sun}
 \affiliation{
 Department of Physics,
  University of Michigan, Ann Arbor, 
 MI 48109-1040, USA
 }

  \author{Stefano Gonella}
 \affiliation{
Department of Civil, Environmental, and Geo- Engineering, University of Minnesota, Minneapolis, MN 55455, USA
 }

  \author{Xiaoming Mao}

 \affiliation{
 Department of Physics,
  University of Michigan, Ann Arbor, 
 MI 48109-1040, USA
 }

\begin{abstract}
Recent progress in topological mechanics have revealed a family of Maxwell lattices that exhibit topologically protected floppy edge modes. These modes lead to a strongly asymmetric elastic wave response. In this paper, we show how topological Maxwell lattices can be used to realize non-reciprocal transmission of elastic waves. 
Our design leverages the asymmetry associated with the availability of topological floppy edge modes and the geometric nonlinearity built in the mechanical systems response to achieve the desired non-reciprocal behavior, which can be further turned into strongly one-way phonon transport via the addition of on-site pinning potentials. Moreover, we show that the non-reciprocal wave transmission can be switched on and off via topological phase transitions, paving the way to the design of cellular metamaterials that can serve as tunable topologically protected phonon diodes.
\end{abstract}

\maketitle
\section{Introduction} 
Over the past decade, significant progress on the development of mechanical analogs of topological states of matter has fueled the new field of ``topological mechanics"~\cite{susstrunk2015observation,kane2014topological,pal2016helical,mousavi2015topologically,prodan2009topological}. A plethora of applications of topological mechanical metamaterials have been proposed, such as uni-directional transport using chiral edge modes~\cite{nash2015topological, mitchell2018amorphous, mitchell2018tunable, mousavi2015topologically, zhang2017experimental, dai2018observation, guo2017topological, susstrunk2015observation, yang2015topological}, transformable topological materials\cite{rocklin2017transformable}, structures with programmed buckling or fracturing patterns\cite{zhang2018fracturing, paulose2015selective}, mechanical laser\cite{abbaszadeh2017sonic}, aperiodic topological metamaterials\cite{zhou2018, zhou2019topological}, geared topological metamaterials\cite{meeussen2016geared}, and dislocation-localized softness\cite{paulose2015topological}. 

A particularly interesting potential application of topological metamaterials is to obtain a \textit{phonon diode}, i.e., a device that only allows sound transmission in one direction.  The main requirement to achieve this goal is to break \textit{reciprocity}. Within linear elasticity, systems with time-reversal symmetry exhibit reciprocity\cite{maznev2013reciprocity}. 
According to Maxwell-Betti's theorem\cite{maxwell1864calculation, betti1872teoria, charlton1960historical}, reciprocity implies that $u_{B}^{j,(1)}/F_{A}^{i} = u_{A}^{i,(1)}/F_{B}^{j}$, where $i, j=x,y,z$ are Cartesian components, $F_A^{i}$ is the $i$-th component of the external force exerted at input point $A$ and $u_B^{j, (1)}$ is the $j$-th component of the linear elastic response probed at output point $B$. In the remainder of this paper, we define the quantity $\chi_{{\rm out},A}^{(1)} = u_B^{j, (1)}/F_A^{i}$ as the linear transmission susceptibility. 
To achieve non-reciprocal transmission one needs to 1) break spatial inversion symmetry and 2) either break time-reversal symmetry, or include nonlinear effects. Major efforts have been devoted to the development of strategies to violate reciprocity by breaking time reversal symmetry. For example, several active metamaterial configurations have been proposed for uni-directional edge wave propagation, such as systems of coupled gyroscopes\cite{wang2015topological, nash2015topological, mitchell2018amorphous, mitchell2018realization}, chiral active fluids and plasma\cite{engheta2005circuit, souslov2019topological}, dynamic phononic lattices\cite{wang2018observation}, spatio-temporally modulated metamaterials\cite{zanjani2014one, trainiti2016non, nassar2017modulated, swinteck2015bulk, fleury2016floquet, salerno2016floquet, chaunsali2016stress, peano2015topological} and active-liquid metamaterials\cite{fleury2014sound, souslov2017topological}.

An alternative route to break reciprocity in mechanical systems consists of leveraging the intrinsic nonlinearity of their elastic response. Recent implementations include nonlinear self-demodulation processes obtained by coupling elastically distinct layers of metamaterials\cite{devaux2015asymmetric, merkel2014directional, cebrecos2016asymmetric}, unidirectional guiding of strongly nonlinear transition waves in a bistable lattice\cite{nadkarni2016unidirectional}, static non-reciprocal elastics\cite{coulais2017static}, acoustic switching and rectification\cite{boechler2011bifurcation, liang2010acoustic, liu2015frequency} and broadband acoustic diodes\cite{gu2016broadband}. 
In this paper, we present an approach for non-reciprocal wave transmission in lattice systems, in which the task of breaking space inversion symmetry is accomplished through the activation of topological floppy edge modes, and the nonlinear response requirements are fulfilled by the geometric nonlinearity of the lattice deformation. The main advantage of the proposed design stems from the topological protection of the edge modes, which endows the non-reciprocal phenomena with  robustness against potential defects and disorder.

Maxwell lattices are central-force lattices with average coordination number $\langle z\rangle = 2d$ ($d$ is the spatial dimension), which puts them on the verge of mechanical instability\cite{zhou2018, kane2014topological, rocklin2016mechanical, lubensky2015phonons}. They host topologically protected edge modes at zero frequency (floppy modes) which are governed by the topology of the equilibrium and compatibility matrices and therefore ultimately depend on the lattice geometry\cite{kane2014topological}. The topological edge modes lead to strongly asymmetric edge stiffness, which has been shown to result in asymmetric wave propagation characteristics, whereby certain edges allow waves to propagate into the bulk, and others localize energy at the boundaries\cite{ma2018edge}. 
Despite this asymmetry, the transmission of linear elastic waves is still reciprocal, meaning that the linear transmission susceptibilities $\chi_{{\rm out},A}^{(1)}$ and $\chi_{{\rm out},B}^{(1)}$ between points A and B in space are equal, in accordance with Maxwell-Betti's theorem.

To achieve non-reciprocity, we need to operate the lattice in the nonlinear regime. Here, we pursue this by propagating second harmonics whose generation and intensity are controlled by the amplitude of the excitation. 
The generation of higher harmonics in mechanical systems with multi-modal dispersive behavior and the resulting opportunities for unconventional wave manipulation and functionality enrichment in elastic metamaterials have been the object of a number of recent studies\cite{deng2005experimental,matlack2011experimental,ganesh2015modal,ganesh2017nonlinear,jiao2019doubly,pal2018amplitude,jiao2018mechanics,jiao2018intermodal}.

In this paper we show that second harmonic modes in topological Maxwell lattices are strongly nonreciprocal, due to the contrast in stiffness between floppy and non-floppy edges, which is a topologically protected property. Further, we demonstrate that, by blocking the linear (first harmonic) modes via on-site pinning potentials (which can be realized by placing the lattice on a soft substrate), the system works as a phonon diode, in which transmission (with frequency doubled) is effectively observed in only one direction. 
Finally, we revisit the notion that topological kagome lattices can be reversibly transformed between different topological states with contrasting edge state landscapes through a transformation, known as the ``Guest mode", which involves a soft strain of the whole lattice. As a result, these lattices can be switched between strongly non-reciprocal and nearly reciprocal states through simple reversible operations.

\begin{figure*}[htb]
\centering
\includegraphics[width=1\textwidth]{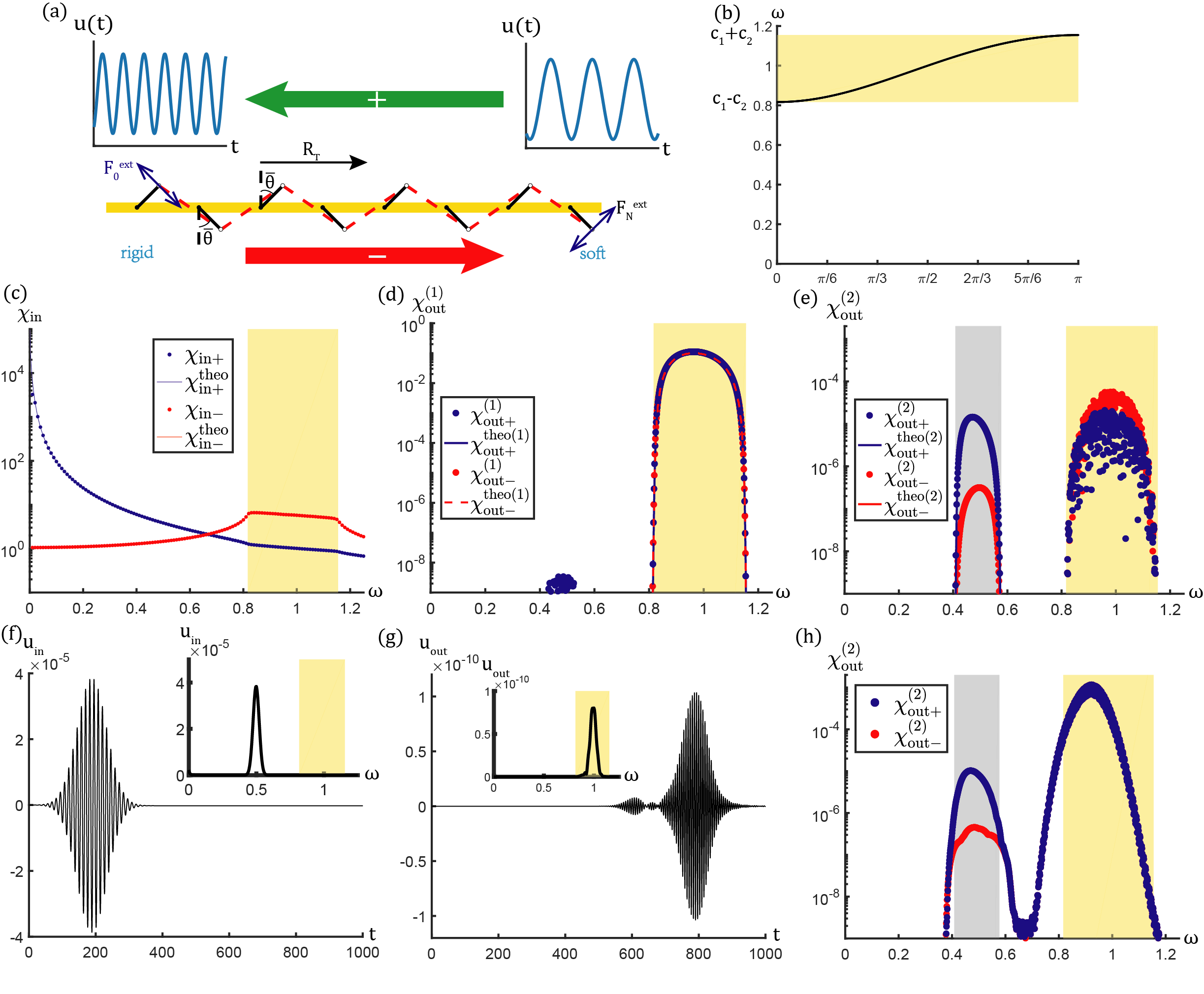}
\caption{Nonreciprocal wave propagation in a 1D nonlinear topological chain. (a) 1D topological mechanical chain\cite{kane2014topological} subjected to open boundary conditions with one floppy mode localized on the right edge. The ``positive (negative)" direction is marked with green (red). We apply a harmonic force $F^{\rm ext}e^{i\omega t}$ with magnitude $F^{\rm ext}=10^{-5}$ and frequency $\frac{1}{2}(c_1-c_2) < \omega = 0.5 < c_1-c_2$ on both edges to excite linear edge modes and second harmonic bulk waves. (b) Band structure $\omega = \omega(ak)$ of the 1D chain. (c) Input-end local response function $\chi_{\rm in}$. We denote the bulk wave region with yellow. (d) Reciprocal transmission of linear waves with $\chi_{\rm out+}^{(1)}=\chi_{\rm out-}^{(1)}$. (e) Non-reciprocal transmission of second harmonic waves with $\chi_{\rm out+}^{(2)}\gg \chi_{\rm out-}^{(2)}$ in the frequency region $\frac{1}{2}(c_1-c_2) < \omega < {\rm min}(\frac{1}{2}(c_1+c_2),c_1-c_2)$ marked in grey. We plot the output second harmonic susceptibility $\chi^{(2)}_{\rm out}(2\omega)$ versus the input driving frequency $\omega$. (f) Input-end response excited by a Guassian tone burst with carrier frequency $\omega = 0.5$. (g) Output-end response featuring carrier frequency $\omega^{(2)} = 2\omega$. (h) Non-reciprocal transmission of second harmonic driven by Guassian tone burst. }\label{fig1}
\end{figure*}

\section{Non-reciprocity in 1D topological mechanical chain}
We start our discussion by revisiting the 1D topological mechanical chain introduced in\cite{kane2014topological}, as shown in fig.\ref{fig1}(a). This is the simplest lattice with topologically protected floppy edge modes that give rise to contrasting boundary rigidity. The chain consists of rigid rotors connected to fixed pivot points separated by lattice constant $a$. The pivot points as well as the rotors are labeled from 0 to $N$. The other ends of the rotors feature particles of mass $m$, and neighboring particles are connected with harmonic springs. The chain is subjected to open boundary conditions (OBC) at rotors $0$ and $N$. The equilibrium configuration is such that rotors form an angle $\bar{\theta}$ relative to the upward and downward normals. The angular displacements are denoted as $\textbf{u}=(r\delta\theta_0,r\delta\theta_1,...,r\delta\theta_N)$, where $\delta\theta_n=\theta_n-\bar{\theta}$. The system consists of $N+1$ degrees of freedom and $N$ constraints, leaving only one topological floppy mode localized on the right boundary. 

Now we imagine driving the chain by a monochromatic harmonic force $F_{g}^{\rm ext}(t) = Fe^{i\omega t}$ applied at the left (right) end, i.e., on rotor $g=0$ ($g=N$), while $F_n^{\rm ext}(t) = 0$ elsewhere. $F$ is assumed small enough that $\delta\theta_n \ll 1$, $\forall n = 0,...,N$, which validates perturbation theory. We denote $\textbf{F}_g^{\rm ext} = (F_0, F_1, ...,F_N)$ as the array of external forces and, as we mentioned previously, $\textbf{F}_0^{\rm ext} = (F, 0, ...,0)$ and $\textbf{F}_N^{\rm ext} = (0, 0, ...,F)$. By expanding $\textbf{u} = \textbf{u}^{(1)}+\textbf{u}^{(2)}+\mathcal{O}(F^3)$ in orders of $F$, we can solve for the linear elastic mode $\textbf{u}^{(1)}$ and for second harmonic mode $\textbf{u}^{(2)}$, respectively. 

We define the input linear response function as $\chi_{\rm in} = |u_{\rm in}^{(1)}|/F_{\rm in}$, where $u^{(1)}_{\rm in}$ is the linear displacement of the rotor that is being driven. We also define $\chi_{\rm out}^{(1)} = |u_{\rm out}^{(1)}|/F_{\rm in}$ ($\chi_{\rm out}^{(2)} = |u_{\rm out}^{(2)}|/F_{\rm in}$), where $u^{(1)}_{\rm out}$ ($u^{(2)}_{\rm out}$) is the linear displacement (second harmonic displacement) at the boundary rotor opposite to the driven side. 

To the linear order, Newton's equation of motion is 
\begin{eqnarray}\label{eom1}
 m\ddot{\textbf{u}}_g^{(1)} = 
\textbf{F}_g^{\rm ext}-\textbf{D}\textbf{u}_g^{(1)}
-\eta \dot{\textbf{u}}_g^{(1)},
\end{eqnarray}
where $\eta$ is the damping coefficient, $m$ is the particle mass, and the lower index $g$ indicates that the force is applied at the left end if $g=0$ (right end if $g=N$). The dynamical matrix is $\textbf{D}=K\textbf{C}^T\textbf{C}$, where $\textbf{C}_{ij}=c_1\delta_{ij}-c_2\delta_{i,j+1}$ is the compatibility matrix, and $c_1$, $c_2$ are coefficients determined by lattice geometry (see SI.\uppercase\expandafter{\romannumeral2}). In the static limit, the linear elasticity is captured by the compatibility matrix $\textbf{C}$: the floppy mode is localized on the right (left) end if $|c_1|>|c_2|$ ($|c_1|<|c_2|$). The topological protection of this floppy mode arises from the winding number of the compatibility matrix in the complex plane, and is therefore invariant against continuous deformations to the geometry of this 1D chain unless the gap closes\cite{kane2014topological}. Without losing generality, in the rest of this paper we always let $c_1>c_2>0$ by allowing rotors to tilt rightwards with $\bar{\theta}>0$, so the edge floppy mode is localized on the right end of the chain. Following this convention, it is convenient  to denote $\chi_{\rm in+} = |u_N^{(1)}|/F_N$, $\chi_{\rm in-} = |u_{0}^{(1)}|/F_0$, and $\chi_{\rm out+}^{(1,2)} = |u_0^{(1,2)}|/F_N$, $\chi_{\rm out-}^{(1,2)} = |u_N^{(1,2)}|/F_0$, where $+$ ($-$) indicates that external driving is applied at the soft (rigid) end.

\begin{figure*}[htb]
\centering
\includegraphics[width=1\textwidth]{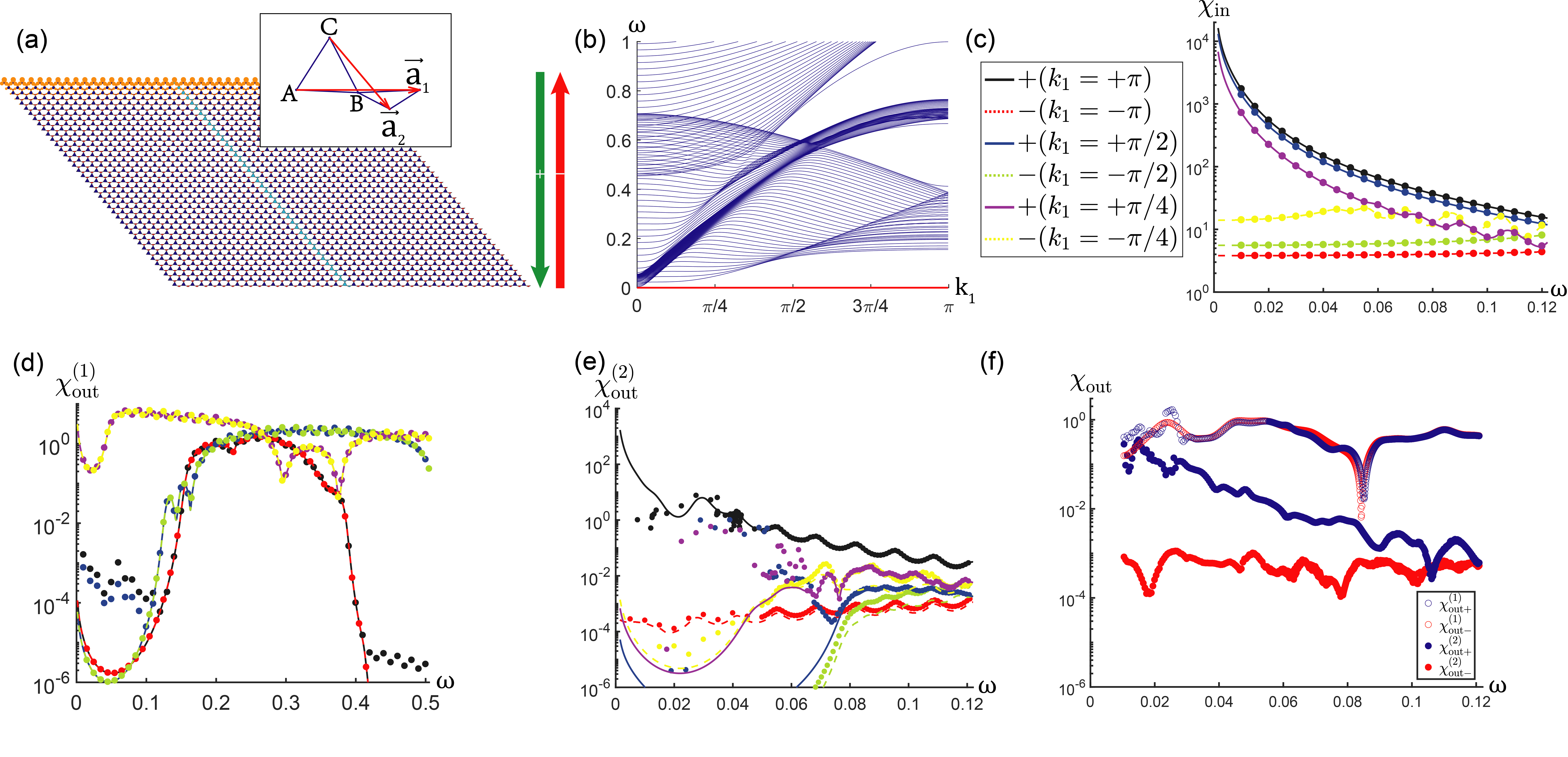}
\caption{Nonreciprocal wave propagation in a 2D nonlinear topological kagome lattice. (a) Topological kagome lattice, with unit cell shown in the inset, where A, B and C label the sites and ($\vec a_1$, $\vec a_2$) are the primitive vectors. The floppy modes are localized on the top  boundary. Total floppy modes amplitude on each site is shown by the size of the orange disks on the site. The lattice consists of $40 \times (40-1)$ unit cells and an additional layer of $40$ C-sites at the bottom boundary. The lattice is subjected to periodic boundary condition in $\vec a_1$ and open boundary conditions on the top and bottom sides. 
The supercell strip used in our analysis is marked with cyan. (b) Supercell band structure, where $k_1=\vec k\cdot \vec a_1$. The dispersion branch of topological floppy modes is marked with red. In (c), (d), (e) and (f) we employ both of analytic calculations (curves) and Newtonian Mechanics simulations (dots) to measure the input local response function $\chi_{\rm in}$ and the output linear (second harmonic) susceptibilities $\chi_{\rm out}^{(1)}$ ($\chi_{\rm out}^{(2)}$). (c) Strongly asymmetric edge response function $\chi_{\rm in}(\omega, k_1)$ with $\chi_{\rm in+}(\omega, k_1)\gg \chi_{\rm in-}(\omega, -k_1)$ for waves below the bulk band ($\omega \lesssim 0.06$). The lattice is driven by a monochromatic excitation force which is spatially periodic in $\vec a_1$, with the wave number $k_1$ for the top boundary ($-k_1$ for the bottom). 
(d) Reciprocal transmission curves of linear elastic waves with $\chi^{(1)}_{\rm out+}(k_1)=\chi^{(1)}_{\rm out-}(-k_1)$, where the external driving force amplitude $F^{\rm ext}=10^{-8}$. 
(e) Non-reciprocal transmission curves of second harmonic waves with $\chi_{\rm out+}^{(2)}(2\omega, 2k_1)\gg \chi_{\rm out-}^{(2)}(2\omega, -2k_1)$, where the external force amplitude $F^{\rm ext}=10^{-4}$. (f) Non-reciprocal transmission of linear and second harmonic modes with monochromatic point driving force. The transmission of second harmonic waves is less non-reciprocal compared to (e) since the point shaking force has all $k_1$ wave number components. The first harmonic modes are bulk modes since long wavelength components are in the band.}\label{fig2}
\end{figure*}

Above zero frequency, the dispersion relation of the bulk phonon mode reads $\omega=[(c_1-c_2)^2+4c_1c_2\sin^2\left(\frac{1}{2}ak\right)]^{1/2}$, where $k$ is the wave number. Linear elastic waves driven by external excitations with $c_1-c_2 < \omega < c_1+c_2$ are bulk modes. $\chi_{\rm in\,+}$ and $\chi_{\rm in\,-}$ are at the same order of magnitude. As $\omega$ falls below $c_1-c_2$, linear modes localize on lattice boundaries. $\chi_{\rm in\,+}/\chi_{\rm in\,-}$ monotonically increases to infinity as $\omega$ approaches the static limit (see SI.\uppercase\expandafter{\romannumeral2} for details). Although the stiffness differs dramatically (by orders of magnitude) on opposite boundaries, the linear elastic transmission is still reciprocal, meaning that $\chi_{\rm out+}^{(1)} =\chi_{\rm out-}^{(1)}$ as a manifestation of Maxwell-Betti's theorem. We verified this equality both analytically and numerically, as shown in fig.\ref{fig1}(d). 

Interestingly, higher order harmonics with $\omega^{(n)} = n\omega$ that are nonlinearly generated by the edge modes are bulk modes as long as $c_1-c_2 < \omega^{(n)} < c_1+c_2$. In what follows, we study whether these second harmonic modes carry non-reciprocal characteristics. 
The Newton's equation of motion for second harmonic modes is
\begin{eqnarray}\label{eom2}
m\ddot{\textbf{u}}_g^{(2)} = \textbf{f}^{(2)}(\textbf{u}_g^{(1)})-\textbf{D}\textbf{u}_g^{(2)}-\eta \dot{\textbf{u}}_g^{(2)},
\end{eqnarray}
where $\textbf{f}^{(2)}(\textbf{u}_g^{(1)})$ is the second harmonic effective driving force generated by the linear displacement $\textbf{u}_g^{(1)}$, as defined in Eq.(\ref{eom1}) (see SI.\uppercase\expandafter{\romannumeral2} for details). Since the effective driving is quadratic in $\textbf{u}^{(1)}_g$, it triggers second harmonic modes with amplitude $|u_n^{(2)}|\propto |u_{\rm in}^{(1)}|^2$ and frequency $2\omega$. External excitations with $\frac{1}{2}(c_1-c_2)<\omega<{\rm min\,}(\frac{1}{2}(c_1+c_2), c_1-c_2)$ excite linear edge modes as well as second harmonic bulk waves. For a given magnitude of excitation $F$, the input-end linear response measured at the right edge is far greater than its counterpart measured at the left edge ($\chi_{\rm in+}\gg \chi_{\rm in-}$). As a result, the global wave amplitude experienced by the chain is much greater when the chain is driven from the right, leading also, in return, to significantly stronger second harmonics generation, i.e., $|\textbf{u}_{ g=N}^{(2)}|\gg |\textbf{u}_{g=0}^{(2)}|$. 
Through analytical and numerical calculations we can show that $\chi_{\rm out+}^{(2)}\gg \chi_{\rm out-}^{(2)}$, meaning that the transmission of second harmonics is non-reciprocal, as reported in fig.\ref{fig1}(e). This non-reciprocity result can be generalized to the $n$-th harmonic mode: we obtain that $\chi_{\rm out+}^{(n)}\gg\chi_{\rm out-}^{(n)}$ if $\frac{1}{n}(c_1-c_2)<\omega<{\rm min\,}(\frac{1}{n}(c_1+c_2), c_1-c_2)$. We note that, besides the low-frequency regime $\omega < c_1-c_2$, linear modes with high frequencies $\omega > c_1+c_2$ can also localize on edges. However, they are not of interest in this paper, because the associated nonlinear harmonics are also edge excitations which cannot propagate across the lattice and therefore cannot contribute to transmission. 

It is interesting to ask whether non-reciprocity still holds if the monochromatic harmonic excitation is replaced by a tone burst excitation with carrier frequency $\omega$ and Gaussian amplitude modulation, having the form $F(t)\sim F e^{i\omega t - (t-t_0)^2/\tau^2}$, where  the parameter $\tau$ controls the spread of the Gaussian and $t_0$ denotes the trigger time of the packet. Since, in Fourier space, the input signal is a Gaussian function with full width at half maximum $\Delta\omega = 2\sqrt{\ln 2}/\omega \tau$, we expect that the transmission of nonlinear modes is still non-reciprocal. This conjecture is verified by numerical analysis as shown in fig.\ref{fig1}(g).

It is important to note that the key ingredient to achieve non-reciprocity  is the contrast in rigidity between opposite edges, and not the topological protection of the edge modes. In principle, any system with asymmetric boundary stiffness, whether this is topological or not, can realize non-reciprocity if such asymmetry is used in conjunction with nonlinear elasticity\cite{coulais2017static}. However, topologically protected floppy modes enjoy the additional attribute of being robust against disorder, noise, and stochastic damage. More interestingly, topological lattices are switchable, meaning that the topological polarization can be changed via simple, reversible operations that modify their geometry.  For example, the 1D chain discussed above can be flipped to have the opposite topological polarization by propagating a soliton through the chain\cite{chen2014nonlinear}. As we shall discuss in section \uppercase\expandafter{\romannumeral4}, 2D topological kagome lattices can undergo a geometric change through a soft strain of the whole lattice, called the ``Guest mode"\cite{guest2003determinacy}, to manipulate topological phases, control floppy mode localization\cite{rocklin2017transformable} and thus boundary stiffness. Consequently, the transmission of nonlinear waves can be switched from non-reciprocal to reciprocal by reconfiguring the lattices from their topological to their non-topological form.

\section{Non-reciprocity in topological kagome lattice}
Having established non-reciprocity for a 1D topological chain, we now ask if the same is true for a 2D topological lattice\cite{kane2014topological}. To this end, we consider the topological kagome lattice shown in fig.\ref{fig2}(a). The lattice is ideal, i.e., it consists of point masses connected by nearest-neighbor linear springs. 
The unit cell contains one equilateral and one isosceles triangles, which are constructed from 6 bonds and 3 nodes marked by A, B and C. The side length of the equilateral triangle and the longer edge of the isosceles triangle are $l_0$, while the shorter edge is $l_0/\sqrt{3}$. The twist angle of the isosceles triangle is $5^\circ$ counterclockwise, which makes the longer edge of isosceles triangle inclined by $5^\circ$ relative to the bottom edge of equilateral triangle (marked $\overline{\rm AB}$ in fig.\ref{fig2}(a)). $\vec a_1$ and $\vec a_2$ are the lattice primitive vectors. The lattice, spanning the area $|N_1\vec a_1\times (N_2-1)\vec a_2|$ in real space, is composed of $N_1\times (N_2-1)$ unit cells and an additional layer of $C$-sites at the bottom edge of the lattice to complete the triangles. It is subjected to periodic boundary condition in $\vec a_1$ and open boundary conditions at the top and bottom edges.

We start by introducing a \textit{supercell} analysis of this lattice. For convenience we denote $k_{1} = \vec k\cdot \vec a_{1}$ and $k_{2} = \vec k\cdot \vec a_{2}$ as the wavenumbers along the primitive vectors. We further decompose the lattice in supercell strips directed along $\vec a_2$, as marked in cyan in fig.\ref{fig2}(a), and we apply Bloch's conditions along $\vec a_1$. 
Within a supercell, the unit cells are labeled from 1 to $N_2-1$ going from top to bottom, and the C-sites on the bottom layer are labeled as $N_2$. 
The internal nodal displacements of unit cell $n_2$ (with $1\le n_2 \le N_2-1$) are denoted as $\textbf{u}_{n_2} = (u_{n_2 A}^x,u_{n_2 A}^y,u_{n_2 B}^x,u_{n_2 B}^y,u_{n_2 C}^x,u_{n_2 C}^y)$. The displacement field of supercell strip is therefore denoted as $\textbf{u}=(\textbf{u}_{1},\textbf{u}_{2},...,\textbf{u}_{N_2})$. The lattice is driven by a monochromatic harmonic force acting vertically and with amplitude varying periodically in the $\vec a_1$ direction, i.e., $\vec F_{n_1, g}^{\rm ext}(t) = e^{i\omega t-ik_1n_1} \vec F_C $ with $\vec F_C = (0,F)$ at the top boundary (bottom boundary) of C-sites labeled by $g=1$ ($g=N_2$) and $\vec F_{n_1, n_2}^{\rm ext}(t) = 0$ otherwise. $F$ is assumed to be small such that all $|\vec u_{n_1n_2,A}|, |\vec u_{n_1n_2,B}|$ and $|\vec u_{n_1n_2,C}| \ll l_0$, validating perturbation theory. By expanding $\textbf{u}=\textbf{u}^{(1)}+\textbf{u}^{(2)}+\mathcal{O}(F^3)$, we can solve Newton's equation of motion for the linear mode $\textbf{u}^{(1)}$ and for second harmonic mode $\textbf{u}^{(2)}$, as detailed in the SI.

The analysis of wave propagation in the 2D lattice follows the steps used for the 1D chain, albeit with the additional wavenumber $k_1$ describing spatial variation in the horizontal direction (with periodic boundary conditions).  Specifically, the dynamical matrix of this super cell strip is $\textbf{D} = K\textbf{C}^\dag(k_1)\textbf{C}(k_1)$. $\textbf{C}(k_1)$ is the compatibility matrix given by $\textbf{C}_{ij}(k_1) = \textbf{C}_1(k_1)\delta_{ij}+\textbf{C}_2(k_1)\delta_{i+1,j}$, where $\textbf{C}_1(k_1)$ and $\textbf{C}_2(k_1)$ are intra-cell and inter-cell compatibility matrices, respectively (see SI.\uppercase\expandafter{\romannumeral3} for details). To the linear order of displacement, Newton's equation of motion is the same as Eq.(\ref{eom1}), where $\eta$ and $m$ are damping coefficient and particle mass, respectively, and $g$ indicates that the input force is applied at the layer of C-sites indexed $g=1$ at the top (layer of C-sites indexed $g=N_2$ at the bottom) of the lattice. The static system is characterized by the \textit{polarization vector} $\vec R_T$, which is a topological invariant.
Mechanical lattices with a well-defined polarization exhibit topological floppy edge modes exponentially localized at the boundary towards which $\vec R_T$ points. The configuration of fig.\ref{fig2}(a) has a polarization vector $\vec R_T = \vec a_1-\vec a_2$. The floppy modes are therefore localized on the top edge, making this edge much softer than the bottom one. We use lower index $+$ ($-$) to indicate that the external signal is applied where the floppy modes are localized (opposite to the floppy mode localization). It is therefore convenient to denote $\chi_{\rm in+} = |u_{1}^{y, (1)}|/F_{1}^y$ and $\chi_{\rm in-} = |u_{N_2}^{y, (1)}|/F_{N_2}^y$ as the soft edge and rigid edge linear response functions, respectively. Similarly, we denote $\chi_{\rm out+}^{(1)} = |u_{N_2}^{y, (1)}|/F_{1}^y$ ($\chi_{\rm out+}^{(2)} = |\vec u_{N_2}^{(2)}|/F_{1}^y$) and $\chi_{\rm out-}^{(1)} = |u_{N_2}^{y, (1)}|/F_{1}^y$ ($\chi_{\rm out-}^{(2)} = |\vec u_{N_2}^{(2)}|/F_{1}^y$) as the linear (second harmonic) transmission susceptibilities driven by external forces applied at the soft and rigid boundaries, respectively. 

Linear wave propagation is governed by the supercell band structure which stems from the eigenvalue problem $\det(\textbf{D}-m\omega^2 \textbf{I}) = 0$. The band structure is gapped except for the trivial translational zero mode point at $k_1=0$. Given the wave number $k_1$ of the applied force, the linear response is a bulk mode if $\omega > \Delta(k_1)$, where $\Delta(k_1)$ is the lowest bulk eigenvalue in the band structure, and $\chi_{\rm in+}$ and $\chi_{\rm in-}$ are of the same order of magnitude. As $\omega$ falls below $\Delta(k_1)$, linear modes localize on the soft boundary of the lattice. $\chi_{\rm in\,+}/\chi_{\rm in\,-}$ monotonically increases to infinity as $\omega$ approaches zero. Despite the contrasting boundary stiffness at low-frequencies, the linear elastic transmission is still reciprocal, i.e., $\chi_{\rm out+}^{(1)}(k_1) =\chi_{\rm out-}^{(1)}(-k_1)$, similar to what we obtained for the 1D topological chain. We validate this equality through analytical and numerical calculations, as shown in fig.\ref{fig2}(d).

While linear modes can localize on the lattice boundaries, nonlinearly generated components with $(\omega^{(n)}, k_1^{(n)}) = (n\omega, n k_1)$ can be bulk waves as long as $\omega^{(n)} > \Delta(k_1^{(n)})$. 
The equation of motion for the second harmonic mode is given by Eq.(\ref{eom2}). External excitations with frequency $\frac{1}{2}\Delta(2k_1) < \omega < \Delta(k_1)$ generate linear boundary modes and second harmonic bulk waves. Moreover, given the same magnitude $F$ of external force, the input-end frequency response function of the floppy edge is far greater than that of the hard edge ($\chi_{\rm in+}\gg\chi_{\rm in-}$), which renders second harmonic bulk modes excited 
at the floppy edge much greater than those excited at the hard edge, i.e., $|\textbf{u}_{ g=N_2}^{(2)}|\gg |\textbf{u}_{g=1}^{(2)}|$. The transmission of second harmonics is therefore strongly non-reciprocal, with $\chi_{\rm out+}^{(2)}\gg \chi_{\rm out-}^{(2)}$, which is verified numerically as shown in fig.\ref{fig2}(e). This conclusion can be generalized to the non-reciprocal transmission of $n$-th harmonic mode with $\chi_{\rm out+}^{(n)}\gg \chi_{\rm out-}^{(n)}$ if $\frac{1}{n}\Delta(nk_1) < \omega < \Delta(k_1)$.

\begin{figure*}[htb]
\centering
\includegraphics[width=1\textwidth]{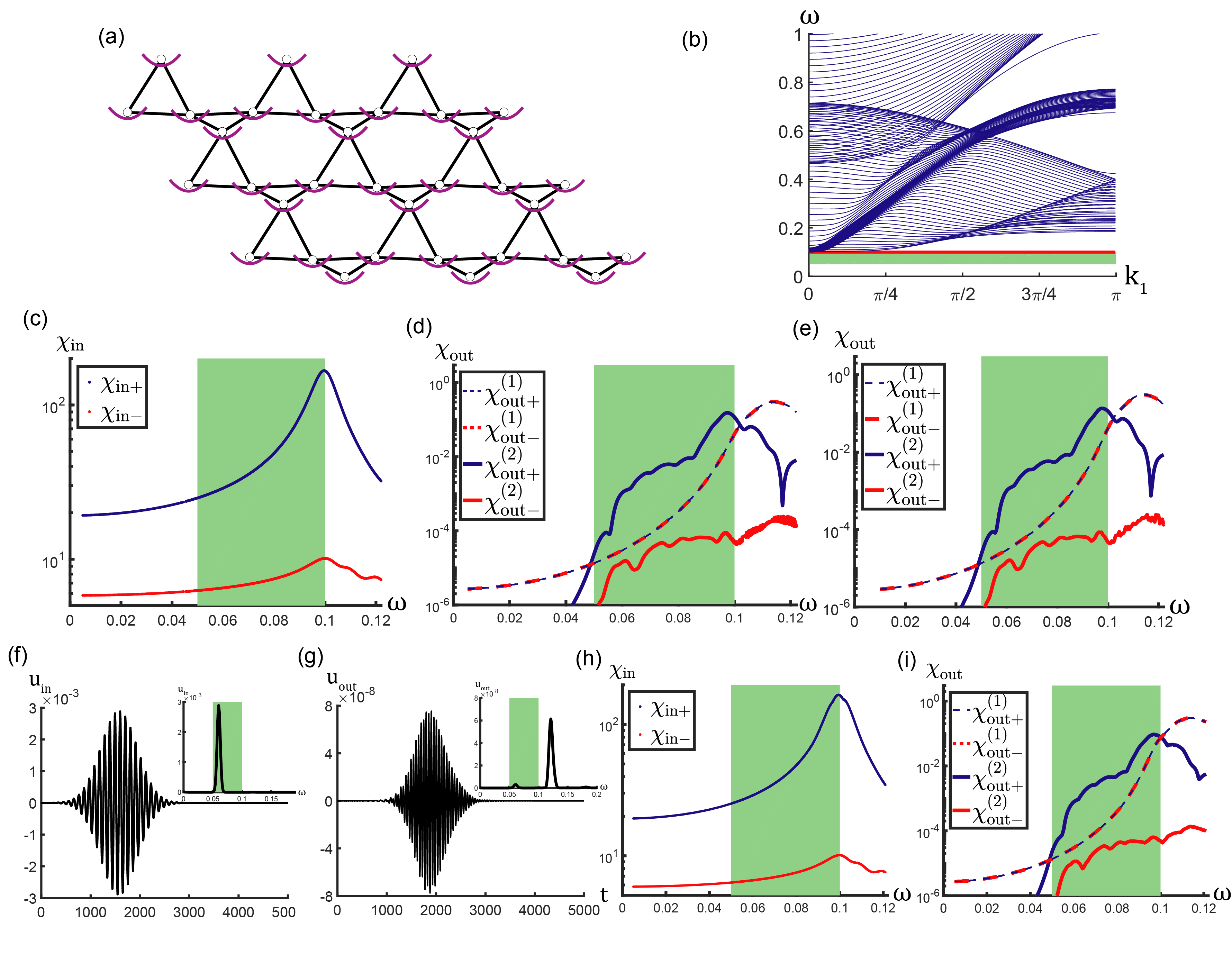}
\caption{One-way propagation of second harmonic waves in a topological kagome lattices with on-site pinning potentials. (a) Schematic illustration of the lattice with on-site potential. (b) Band structure of supercell strip with on-site potential $K'=K/100$, which fully gapped the spectrum at low frequency. We mark the edge mode excitation with red at $\omega = \sqrt{K'/m}$, and the frequency range where second harmonics $2\omega$ are in the band with green. In (c), (d), (e), (h) and (i) we employ Newtonian Mechanics simulations to measure the input local response function $\chi_{\rm in}$ and the output susceptibility $\chi_{\rm out}$ against point shaking force at an arbitrary C-site on top or bottom. (c) Asymmetric stiffness of the boundary at which the point harmonic excitation is applied (with force amplitude $f=10^{-4}$). (d) Non-reciprocal transmission of second harmonic modes. The transmission susceptibility in the positive direction (i.e., transmission from soft edge to hard edge), marked in blue, is much larger than that in the negative direction marked in red, and also much larger than the first harmonics (dashed lines). (e) Non-reciprocal transmission of second harmonic modes calculated including bending stiffness $\kappa = 10^{-5}K$. (f) Input-end displacement time history for monochromatic point excitation in the form of Gaussian tone burst (frequency spectrum in the inset). (g) Output-end displacement time history for tone burst excitation where the frequency is twice of the input wave frequency. (h) Input-end frequency response for tone burst excitation. (i) Output linear and second harmonic transmission susceptibilities for tone burst excitation. The result is very similar to fig.\ref{fig4}(d), confirming the robustness of the results in transitioning from steady-state to transient regimes of excitation.
}\label{fig4}
\end{figure*}

While, so far, the analysis has followed almost \textit{verbatim} the same steps of the 1D problem, one important difference is that the 2D lattice phonon band depends on $k_1$ (the wave number in the horizontal direction imposed along the boundary).  Thus, the width of the gap $\Delta$ and the resulting availability of nonreciprocal propagation depend on the choice of $k_1$.  In particular, the lattice always has translational zero modes since $\lim_{k_1\to 0}\Delta(k_1) \to 0$. As a result, if we drive the system with a point force applied at a given location on the boundary (which ostensibly excites all values of $k_1$), we are bound to observe weaker signatures of non-reciprocity. In other words, the differences in behavior observed by exciting the soft and hard edges will be vastly reduced, as the long wavelength components of the excited linear waves are in both cases bulk modes that do not display asymmetry. Moreover, despite the strong non-reciprocity, this kagome lattice cannot be, strictly speaking, defined as a proper phonon diode. This is because the linear mode, which is reciprocal and always transmitted both ways, is much stronger than the second harmonic mode and always dominates the total response, completely overshadowing any asymmetry in the nonlinear response. In order to mitigate the aforementioned issues, in the next section we propose an evolution of the lattice design meant to work as a proper phonon diode for all wavenumbers.

\begin{figure*}[htb]
\centering
\includegraphics[width=1\textwidth]{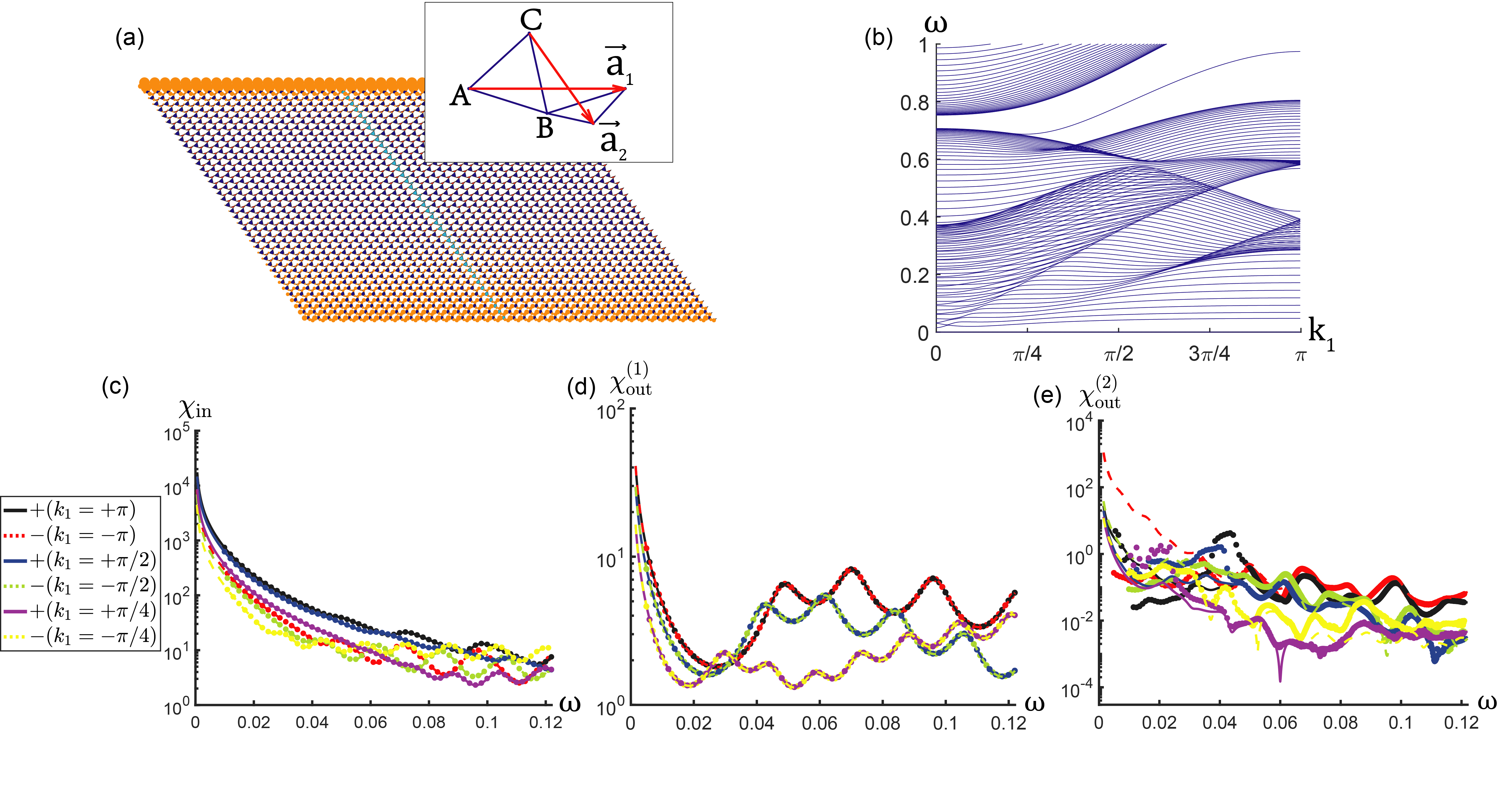}
\caption{Near-reciprocal wave propagation in non-topological kagome lattice, which is related to the topological kagome lattice by a soft-strain reconfiguration. (a) Kagome lattice in the non-topological regime. The floppy modes amplitude shown by the size of the orange disks on each site are localized on both the top and the bottom boundaries. (b) Supercell band structure of non-topological kagome lattice. (c) Different from the topological kagome lattice, the edge response functions $\chi_{\rm in}(\omega, k_1)$ are significantly more symmetric between top and bottom boundaries. We drive the lattice with spatially periodic harmonic force with wave numbers $k_1 = +\pi, +\pi/2, +\pi/4$ on top boundary ($k_1 = -\pi, -\pi/2, -\pi/4$ on bottom). The numerical results are represented by the dots and the theoretical curves are given by solid lines for the top local response functions (dashed lines for the bottom ones). (d) Reciprocal transmission of linear elastic waves in non-topological kagome lattice with external driving amplitude $f=10^{-8}$. (e) Almost reciprocal transmission of second harmonic waves with various wavenumbers. The second harmonic transmission $\chi_{\rm out+}^{(2)}(2\omega, 2k_1)$ is always comparable to $\chi_{\rm out-}^{(2)}(2\omega, -2k_1)$.}
\label{fig3}
\end{figure*}

\section{Non-reciprocity in kagome lattice with on-site potential}
We consider a modification of the topological kagome lattice discussed in section III, where a weak on-site pinning potential is added to every mass point $i$, with $V_i = \frac{1}{2}K' \vec u_i^2$ and $K'\ll K$. This operation, which practically elastically connects each site to a fixed ground point, can be thought of as the equivalent model of placing the lattice on a soft substrate (or soft elastic foundation). It penalizes particles from moving away from their rest positions, and therefore eliminates the lattice trivial translational zero modes. Since the weak pinning rigidly shifts the band up by $\Delta' =  \sqrt{K'/m}$, now the signals with frequency below $\Delta'$ will excite edge modes. 
The weak pinning does not change the landscape of asymmetric boundary stiffness of the lattice, meaning that $\chi_{\rm in+}\gg \chi_{\rm in-}$ still holds as long as $\omega < \Delta'$. Thus, linear edge modes are still preferentially localized on the soft boundary.

The fundamental consequence of having a full low-frequency gap is that, as long as $\frac{1}{2}\Delta'<\omega<\Delta'$, an external excitation that is periodic along the boundary will excite linear edge modes and second harmonic bulk modes for any arbitrary wavenumber $k_1$. Consequently, the non-reciprocal behavior will be observed in the response to a point excitation prescribed at a given location on the boundary, thus eliminating the limitation of the previous configuration. As shown in fig.\ref{fig4}(d), the second harmonic positive transmission is indeed stronger than the linear transmission for a point excitation. We can conclude that the topological lattice with pinning potential is now a well-defined phonon diode. 

These results still hold when finite bending stiffness at the hinges is included\cite{ma2018edge}. In fig.\ref{fig4}(e) our numerical results indeed show that, when bending stiffness is introduced by adding the contribution of next-nearest-neighbor (NNN) interactions, the non-reciprocal transmission is still significant. Finally, the results also hold for a Gaussian tone burst excitation $F(t)\sim Fe^{i \omega t-(t-t_0)^2/\tau^2}$. The numerical analysis results reported in fig.\ref{fig4}(i) show that the transmission is still non-reciprocal, similar as the result of 1D topological mechanical chain.

\section{Lattice reconfiguration and non-reciprocity switching}
An interesting feature of Maxwell lattices with spring-mass interactions is that they can undergo uniform soft deformations, in which all the unit cells are twisted in the same fashion while leaving the bond lengths unstretched. Such uniform deformation, known as the ``Guest mode", can manipulate the geometrical parameters that controls the topological phase of the kagome lattice and, consequently, its polarization and the rigidity established on opposite boundaries\cite{rocklin2017transformable}. Starting from the unit cell configuration in fig.\ref{fig2}(a), by uniformly rotating all the isosceles triangles counterclockwise by $30^\circ$ relative to the hinges on the equilateral triangles, the lattice enters a non-topological phase, as shown in fig.\ref{fig3}(a). The total number of floppy modes remains the same, but, instead of being all localized on the top edge, they localize on both lattice boundaries with nearly comparable stiffness. Thus, given an external force excitation with amplitude $F$ and frequency $\frac{1}{2}\Delta' < \omega < \Delta'$, the transmission of any nonlinearly generated second harmonics is reciprocal, because the linear modes driven from opposite lattice boundaries, which are ultimately responsible for second harmonic generation, have the same order of magnitude. In conclusion, through uniform soft twisting modes that allow reversible reconfiguration between topological and non-topological phases, Maxwell lattices have the ability to switch between reciprocal and non-reciprocal transmission regimes of nonlinear elastic waves without the need to physically disassemble and reassemble the system.

\section{Discussion and Concluding Remarks} 
In this paper we have studied the connection between non-reciprocity and topology in Maxwell lattices. Here, the conditions required for the establishment of non-reciprocal behavior come from the interplay between two factors: on one hand, the availability of floppy edge modes, which yield large boundary deformations and trigger a nonlinear response; on the other hand, the topological polarization, which guarantees asymmetry across the lattice.

Different from the previous work by Coulais \emph{et. al.}\cite{coulais2017static}, who studied static non-reciprocal elasticity in both of topological and non-topological quasi 1-dimensional (1D) mechanical metamaterials, our design focuses on non-zero frequencies. The concept is developed first using a 1D topological mechanical chain and subsequently generalized to 2D topological kagome lattices\cite{kane2014topological}, paving the way to applications in realistic mechanical metamaterials. 
The foundational argument of the proposed concept is that topological floppy edge modes produce contrasting stiffness on opposite lattice boundaries. We found that signals with frequency $\omega < \Delta$, where $\Delta$ is the  onset of a (partial or total) bandgap, excite edge modes. These modes localize asymmetrically, leading to larger deformation that promotes second harmonic generation at the floppy edge. Second harmonic contributions are bulk modes if $\omega > \frac{1}{2}\Delta$, and, in these conditions, they can propagate through the medium. Finally, because of the asymmetry mentioned above, such transmission is highly non-reciprocal.

By adding an on-site pinning potential to every particle, the band structure of the topological kagome lattice is shifted up by $\Delta'$. In these conditions, external signals with $\frac{1}{2}\Delta' <\omega < \Delta'$ excite linear modes that remain localized at the edges and second harmonic bulk modes that propagate across the lattice. Hence, the second harmonic positive transmission is greater than the fundamental mode transmission and is therefore not trivially overshadowed by the linear response. With this improved configuration, this effect is found to be true for any external harmonic excitation applied at the edges, regardless of the wavenumber established along the boundary. Consequently, the result also holds for point excitations, which represent realistic conditions in practice. We have concluded that the lattice with on-site pinning potential fulfills all the requirements to be labeled a phonon diode. In addition, one can control the geometry of the Maxwell lattice through a Guest mode to switch between topological and non-topological phases. This lattice reconfiguration allows us to manipulate reciprocal and non-reciprocal transmission of elastic waves without disassembling or reassembling the structure.

The idea of nonlinear bulk waves driven by linear edge modes is not limited to second harmonics. One can observe $n$-th order harmonic bulk modes at the output end as long as the input frequency satisfies the condition $\frac{1}{n} \Delta < \omega < \frac{1}{n-1} \Delta$ ($n\ge 2$), while all lower-order harmonics are localized on the boundaries and cannot propagate. 
A methodological problem to be considered in performing such an extension is associated with the intrinsic limitations of perturbation theory. Since the amplitude of the output signal becomes exponentially small when the order increases, i.e., $A^{(n)}\sim A^n$, higher-order harmonic bulk modes become progressively more difficult to be observed. It would thus be interesting to study these phenomena in regimes of strong nonlinearity which invalidate perturbation theory. This kind of study would likely present new challenges arising from the interplay between topological states of matter and nonlinear chaos dynamical theory. 

Finally, our investigation so far has been limited to 1D and 2D topological Maxwell lattices. An analogous study of non-reciprocal transmission in 3D topological lattices appears to be possible as a natural extension within the proposed framework. This would open the doors to a broader range of engineering applications and will be one of the next directions in our research. 

This work was supported by the National Science Foundation (Grant No. NSF-EFRI-1741618).

\textbf{Author contributions:} D. Z. performed the theoretical analysis and computations. All authors contributed to the formulation of the problem, analysis of results, and preparation of the manuscript.

\appendix

\section{Reciprocity of linear elastic systems with time reversal symmetry}
With time reversal symmetry, the transmission of linear elastic modes is reciprocal, meaning that the transmission susceptibilities from point $A$ to point $B$ and from point $B$ to point $A$ are equal. This is the essence of the Maxwell-Betti's theorem\cite{maxwell1864calculation, betti1872teoria, charlton1960historical}. In this section we verify this theorem by considering a $d$-dimensional general lattice based on spring-mass interactions. Within linear elasticity, the Newton's equation of motion is 
\begin{eqnarray}\label{A0}
m\ddot{\vec u}_n=-\eta \dot{\vec u}_n -\vec \nabla_{n}{ V}+\vec F_n^{\rm ext}(t), 
\end{eqnarray}
where $n$ denotes a lattice site, $\eta$ is the damping coefficient, $V$ is the lattice potential energy, $\vec \nabla_n = \sum_{i=1}^d\hat{e}_i\partial_{u_n^{(i)}}$, and $\vec F_n^{\rm ext}$ is the external driving force. We rewrite the displacement field as a $Nd$-dimensional vector ${\textbf{u}} = (\vec u_1, \vec u_2, ..., \vec u_N)$, and rewrite the external driving as a $Nd$-dimensional vector ${\textbf{F}}^{\rm ext}$. The linear elastic mode can be calculated as 
\begin{eqnarray}\label{A1}
{\textbf{u}}(\omega) = {\textbf{G}}(\omega){\textbf{F}}^{\rm ext}(\omega), 
\end{eqnarray}
where ${\textbf{G}}(\omega) = \left[{\textbf{D}}+(-m\omega^2+i\eta\omega){\textbf{I}}\right]^{-1}$ is the frequency response function, and ${\textbf{D}}$ is the dynamical matrix. By using an orthogonal transformation $\textbf{S}$ that relates $u_n$ to the normal modes $u_\alpha$ through $u_\alpha = \sum_n S_{\alpha n}u_n$, we can express the normal modes as follows, 
\begin{eqnarray}\label{A2}
u_\alpha(\omega) = G_\alpha F_\alpha^{\rm ext}(\omega),
\end{eqnarray}
where $G_\alpha(\omega) = [\epsilon_\alpha+(-m\omega^2+i\eta\omega)]^{-1}$, and $\epsilon_\alpha$ is the $\alpha^{th}$ eigenvalue of the dynamical matrix $\textbf{D}$. We plug in the driving force at point $A$ to calculate the displacement at $B$, with $u_B(\omega)= \sum_\alpha S_{\alpha B}G_\alpha(\omega) S_{\alpha A}F_A^{\rm ext}(\omega)$. Similarly, we plug in the driving force at $B$ to calculate displacement at $A$ with $u_A(\omega)=\sum_\alpha S_{\alpha A}G_\alpha(\omega) S_{\alpha B}F_B^{\rm ext}(\omega)$. It is evident that $u_A(\omega)/F_B^{\rm ext}(\omega) = u_B(\omega)/F_A^{\rm ext}(\omega)$, meaning that the transmission of linear modes is reciprocal in real space in any elastic system with time reversal symmetry.

\section{Analytical calculation of linear and second harmonic modes in 1D topological mechanical chain}
As shown in fig.1(a), the 1D topological mechanical chain consists of rigid bars of length $r$, free to rotate about hinges separated by the distance $a$, creating repeated 2-site unit cell of length $2a$. A mass point $m$ is attached to the end of each bar, and the neighboring ends are connected by harmonic springs with spring constant $K$. The equilibrium configuration is such that each rotor makes an angle $\bar{\theta}$ relative to the upward or downward normals. 
The angular displacement of rotor $n$ is $u_n=r\delta\theta_n$. We label the rotors from 0 to $N$, with open boundary conditions at rotors 0 and $N$. We apply an external angular driving along the tangential direction of rotor $g$ with $F_n^{\rm ext}(t) = Fe^{i\omega t}\delta_{ng}$. $F$ is not large, making angular displacements of all rotors $u_n\ll r$, which further validates perturbation theory. It is convenient to rewrite the external force as $F\to \lambda F$, where $\lambda\ll 1$. The full Newton's equation of motion is
\begin{eqnarray}\label{B1}
m\ddot{u}_{n,g} & = & (\vec T_{n-1} - \vec T_{n})\cdot \hat{t}_n+F_n^{\rm ext}-\eta \dot{u}_{n,g},
\end{eqnarray}
where the lower index $g$ indicates that the elastic mode is in response to the external driving at rotor $g$. The open boundary conditions at rotor 0 and rotor $N$ are given by 
\begin{eqnarray}\label{B2}
 \vec T_{-1} = \vec T_{N} = 0,
\end{eqnarray}
where $\vec T_{n} = T_{n}\hat{n}_{n}$ is the tension in bond $n$ connecting sites $n$ and $n+1$. $\hat{n}_n$ is the unit vector of bond $n$, and $\hat{t}_n$ is the tangential unit vector of rotor $n$. We expand the tangential component of bond tension in orders of $u_n/r$, denoted by $(\vec T_{n-1} - \vec T_{n})\cdot \hat{t}_n = f_{n}^{(1)}+f_{n}^{(2)}+\mathcal{O}(u^3)$. The leading order is
\begin{eqnarray}\label{B3}
f^{(1)}_{n} = K [c_1c_2(u_{n+1}+u_{n-1})-(c_1^2+c_2^2)u_n],
\end{eqnarray}
with 
\begin{eqnarray}\label{B3.1}
c_{1,2}=\frac{(a\pm 2r\sin\bar{\theta})\cos\bar{\theta}}{\sqrt{a^2+4r^2\cos^2\bar{\theta}}},
\end{eqnarray}
where $\bar{\theta}>0$ is assumed in this paper, leading to $c_1>c_2$. The second order term is
\begin{eqnarray}\label{B4}
f^{(2)}_{n}(u_n)  & = & mr^{-1}\left(C_1u_{n-1}^2+C_2u_{n-1}u_n+C_3u_n^2\right)\nonumber \\
 & + & mr^{-1}\left(C_4u_{n+1}^2+C_5u_{n+1}u_n+C_6u_n^2\right),\quad
\end{eqnarray}
with $C_{1,2,3,4,5,6}$ being constants determined by the geometric parameters of the chain. In our calculations, we choose $a=2r$ and $\theta = \pi/4$. The coefficients are given by 
\begin{eqnarray}\label{B5}
 & {} & 2C_1 = C_5 = \frac{K}{4m}(-1+\sqrt{2})\nonumber \\ 
 & {} & 2C_4 = C_2 = \frac{K}{4m}(-1-\sqrt{2})\nonumber \\
 & {} & C_3 = \frac{K}{24m}(5-\sqrt{2})\qquad 
C_6 = \frac{K}{24m}(5+\sqrt{2}).\quad
\end{eqnarray}
We denote $\textbf{u}=(u_0,u_1,...,u_N)$ as the angular displacement of the particles, and expand it in orders of $\lambda$, with $\textbf{u}=\textbf{u}^{(1)}+\textbf{u}^{(2)}+\mathcal{O}(\lambda^3)$, where $\textbf{u}^{(1)}$ and $\textbf{u}^{(2)}$ are linear and second harmonic modes, respectively. We further denote $\textbf{F}^{\rm ext} = (\lambda F,0,...,0)^T$ ($\textbf{F}^{\rm ext} = (0,0,...,\lambda F)$) as the external driving force driven at rotor $g=0$ ($g=N$), and denote $\textbf{f}^{(2)}(\textbf{u}^{(1)}) = (f_0^{(2)},f_1^{(2)},...,f_N^{(2)})$ as the second harmonic effective feedback force generated by the linear elastice modes. By expanding Eq.(\ref{B1}) up to the second order of $\lambda$, one obtains 
\begin{eqnarray}\label{B5.1}
 & {} & m\ddot{\textbf{u}}_g^{(1)}+m\ddot{\textbf{u}}_g^{(2)}+\mathcal{O}(\lambda^3) = 
\left(\textbf{F}^{\rm ext}-\textbf{D}\textbf{u}_g^{(1)}
-\eta \dot{\textbf{u}}_g^{(1)}\right)\nonumber \\
 & {} & +\left(\textbf{f}_g^{(2)}(\textbf{u}_g^{(1)})-\textbf{D}\textbf{u}_g^{(2)}-\eta \dot{\textbf{u}}_g^{(2)}\right)+\mathcal{O}(\lambda^3),
\end{eqnarray}
where $\textbf{D} = K[(c_1^2+c_2^2)\delta_{ij}-c_1c_2(\delta_{i,j+1}+\delta_{i,j-1})]$ is the dynamical matrix. By matching the equations of motion in orders of $\lambda$, we solve for the linear mode 
\begin{eqnarray}\label{B6}
\textbf{u}_{g}^{(1)}(\omega)=\textbf{G}(\omega)\textbf{F}^{\rm ext}(\omega),
\end{eqnarray}
and for the second harmonic mode 
\begin{eqnarray}\label{B7}
\textbf{u}_{g}^{(2)}(2\omega)=\textbf{G}(2\omega)\textbf{f}^{(2)}_g(\textbf{u}_{g}^{(1)}),
\end{eqnarray}
subjected to the open boundary conditions at rotors $0$ and $N$. The frequency response function of the 1D chain is
\begin{eqnarray}\label{B6.1}
\textbf{G}(\omega)=[\textbf{D}+(-m\omega^2+i\eta\omega)\textbf{I}]^{-1}.
\end{eqnarray}


The dispersion relation of the only bulk phonon mode is $\omega(k) = [(c_1-c_2)^2+4c_1c_2\sin^2\left(ak/2\right)]^{1/2}$, where $k$ is the wave number. The band has lower limit $\Delta = |c_1-c_2|$ and upper limit $\Delta'=|c_1+c_2|$. Thus, the linear mode is a bulk mode if $\Delta<\omega<\Delta'$, while it is an edge mode if $\omega<\Delta$. According to Eq.(\ref{B6}), the generic solution for a linear mode is 
\begin{eqnarray}\label{B9}
 & {} & u_{n,g}^{(1)} = a_g\lambda_1^n+A_g\lambda_2^n\qquad 0\le n\le g\nonumber \\
 & {} & u_{n,g}^{(1)} = b_g\lambda_1^n+B_g\lambda_2^n\qquad g\le n\le N,
\end{eqnarray}
where $\lambda_{1,2}$ are given by 
\begin{eqnarray}\label{B10}
 \lambda_{1,2}= \frac{1}{2}\left(-\gamma \pm \sqrt{\gamma^2-4}\right),
\end{eqnarray}
with $\gamma = \frac{\omega^2}{c_1c_2}-\frac{c_1}{c_2}-\frac{c_2}{c_1}-\frac{i\eta\omega}{c_1c_2}$. Through Eq.(\ref{B6}) and Eq.(\ref{B9}) we can solve for $a_g, A_g, b_g, B_g$. 
We let $g=0$ and $g=N$ to obtain the linear modes when the external driving is applied in at rotors $0$ and $N$. Given the definition of local response function, $\chi_{\rm in}(\omega) = |u_{\rm in}^{(1)}(\omega)|/F_{\rm in}(\omega)$ at rotor $0$ (rigid end, $\chi_{{\rm in} -}$) and rotor $N$ (soft end, $\chi_{{\rm in} +}$), the local response functions are given by
\begin{eqnarray}\label{B12}
 & {} & \chi_{{\rm in} +}(\omega) =(mc_1|c_1-c_2\lambda_1|)^{-1}\nonumber \\
 & {} & \chi_{{\rm in} -}(\omega) =(mc_2|c_2-c_1\lambda_1|)^{-1},
\end{eqnarray}
where we have used $\lim_{N\to\infty}|\lambda_2/\lambda_1|^N= 0$ when $\omega < c_1-c_2$. The ratio $\chi_{{\rm in} +}/\chi_{{\rm in} -}$ tells which end of the 1D chain has greater displacement in response to external loading. In the limit of $\eta\to 0$, analytical calculations reveal that 
\begin{eqnarray}\label{B12.1}
\begin{array}{ccc}
{\chi_{{\rm in} +}}/{\chi_{{\rm in} -}} > 1, &\qquad \omega< \omega^*, \\
{\chi_{{\rm in} +}}/{\chi_{{\rm in} -}} < 1, &\qquad \omega^* < \omega < c_1-c_2, 
\end{array}\qquad
\end{eqnarray}
where $\omega^* = (c_1^2-c_2^2)/\sqrt{2(c_1^2+c_2^2)}$. As long as $\omega< \omega^*$, the response linear edge mode of the right end is greater than that of the left side. The linear transmission susceptibility, defined by $\chi_{\rm out}^{(1)}(\omega) = |u_{\rm out}^{(1)}(\omega)|/F(\omega)$, is given by
\begin{eqnarray}\label{B13}
\chi_{\rm out-}^{(1)}(\omega) = \chi_{\rm out +}^{(1)}(\omega) =\frac{1}{m}\frac{\lambda_1^{-N-1}(\lambda_2-\lambda_1)}{\omega^2-i\eta\omega}
\end{eqnarray}
The fact that the linear transmission susceptibilities are equal is testament to the reciprocity of linear waves. 

We further study the second harmonic modes based on Eq.(\ref{B7}). To calculate the second harmonic mode displacement $\textbf{u}_g^{(2)}$, we notice that $\textbf{u}_g^{(2)} = \sum_{g'=0}^N \textbf{u}_{g'g}^{(2)}$, where $\textbf{u}_{g'g}^{(2)}$ is the displacement field of the chain in response to external force $f^{(2)}_{g'g}$ applied at a single rotor $g'$. The displacement field is given by
\begin{eqnarray}\label{B14}
\textbf{u}_{g'g}^{(2)}(2\omega)=\textbf{G}(2\omega)\textbf{f}^{(2)}_{g'g}(2\omega),
\end{eqnarray}
where we denote the external driving at rotor $g'$ as a $N\times 1$ vector $\textbf{f}^{(2)}_{g'g}=(0,0,...,f^{(2)}_{g'g},...,0)$. The generic solution of Eq.(\ref{B14}) is of the following form
\begin{eqnarray}\label{B15}
 & {} & u_{n,g'g}^{(2)} = c_{g'g}\mu_1^n+C_{g'g}\mu_2^n\qquad 0\le n\le g'\nonumber \\
 & {} & u_{n,g'g}^{(2)} = d_{g'g}\mu_1^n+D_{g'g}\mu_2^n\qquad g'\le n\le N,
\end{eqnarray}
where $\mu_{1,2}$ satisfy 
\begin{eqnarray}\label{B16}
\mu_{1,2} = \frac{1}{2}\left(-\nu \pm \sqrt{\nu^2-4}\right),
\end{eqnarray}
with $\nu = \frac{4\omega^2}{c_1c_2}-\frac{c_1}{c_2}-\frac{c_2}{c_1}-\frac{2i\eta\omega}{c_1c_2}$. Through Eq.(\ref{B14}) and Eq.(\ref{B15}) we solve for $c_{g'g}, C_{g'g}, d_{g'g}, D_{g'g}$ to determine $\textbf{u}^{(2)}_{g'g}(2\omega)$. 
Finally, given the definition of second harmonic transmission susceptibility, $\chi_{\rm out}^{(2)} = |u_{\rm out}^{(2)}(2\omega)|/F(\omega)$, we obtain
\begin{eqnarray}\label{B18}
\chi_{\rm out +}^{(2)}
 & = & \frac{1}{F}\sum_{g'=0}^N\left(c_{g',g=N}+C_{g',g=N}\right)\nonumber \\
\chi_{\rm out  -}^{(2)}
 & = & \frac{1}{F}\sum_{g'=0}^N\left(\mu_1^N d_{g',g=0}+ \mu_2^N D_{g',g=0}\right), \quad
\end{eqnarray}
The second harmonic transmission susceptibility is determined by $\textbf{f}_{g'g}^{(2)}$, which in turn is proportional to the square of linear elastic waves amplitude. This consideration is essential in explaining how the asymmetric local response function $\chi_{\rm in+}^{(1)}\gg \chi_{\rm in-}^{(1)}$ results in the non-reciprocal transmission of second harmonic modes ($\chi_{\rm out +}^{(2)}\gg \chi_{\rm out -}^{(2)}$).

\section{Analytical calculation of linear and second harmonic modes in a 2D topological kagome lattice}
In this section we calculate linear and second harmonic modes in a 2D topological kagome lattice. The unit cell is shown in fig.2(a) and consists 6 bonds, with rest lengths $l_i$ and unit vector directions $\hat{n}_i=(\cos\theta_i,\sin\theta_i)$, $i=1,2,...,6$. We define the $2\times 2$ ``dynamical matrix" of bond $i$, as
\begin{eqnarray}\label{C2}
\textbf{D}_i = K\hat{n}_i\hat{n}_i \qquad i = 1,2,...,6
\end{eqnarray}
$\vec a_1$ and $\vec a_2$ are primitive vectors. The lattice has periodic boundary condition in $\vec a_1$ and open boundary condition in $\vec a_2$, leaving the top and bottom boundaries open. 

The compatibility matrix of the quasi-1D strip of deformed kagome lattice is $\textbf{C}_{ij}(k_1) = \textbf{C}_1(k_1)\delta_{ij}+\textbf{C}_2(k_1)\delta_{i+1,j}$, where $\textbf{C}_1(k_1)$ and $\textbf{C}_2(k_1)$ are intra-cell and inter-cell compatibility matrices, respectively:
\widetext
\begin{eqnarray}\label{C2.1}
\textbf{C}_1 = 
\left(
\begin{array}{cccccc}
\cos\theta_1 & \sin\theta_1 & -\cos\theta_1 & -\sin\theta_1 & 0 & 0\\
0 & 0 & \cos\theta_2 & \sin\theta_2 & -\cos\theta_2 & -\sin\theta_2 \\ 
-\cos\theta_3 & -\sin\theta_3 & 0 & 0 & \cos\theta_3 & \sin\theta_3\\
-e^{ik_1}\cos\theta_4 & -e^{ik_1}\sin\theta_4 & 0 & 0 & 0 & 0\\
e^{ik_1}\cos\theta_5 & e^{ik_1}\sin\theta_5 & -\cos\theta_5 & -\sin\theta_5 & 0 & 0 \\
0 & 0 & \cos\theta_6 & \sin\theta_6 & 0 & 0\\
\end{array}
\right),
\textbf{C}_2 = 
\left(
\begin{array}{cccccc}
0 & 0 & 0 & 0 & 0 & 0\\
0 & 0 & 0 & 0 & 0 & 0\\ 
0 & 0 & 0 & 0 & 0 & 0\\
0 & 0 & 0 & 0 & \cos\theta_4 & \sin\theta_4\\
0 & 0 & 0 & 0 & 0 & 0 \\
0 & 0 & 0 & 0 & -\cos\theta_6 & -\sin\theta_6\\
\end{array}
\right).\qquad 
\end{eqnarray}
\endwidetext

We simplify the lattice to a quasi-1D strip by applying Bloch condition along $\vec a_1$ to obtain a finite supercell strip with $N_2$ unit cells. Thus, the elastic wave $\textbf{u}_{n_1n_2} = e^{ik_1n_1}\textbf{u}_{n_2}$, with $k_1 = \vec k\cdot \vec a_1$. We denote the displacement of cell $n_2$ as $\textbf{u}_{n_2}= (u_{n_2 A}^x, u_{n_2 A}^y, u_{n_2 B}^x, u_{n_2 B}^y, u_{n_2 C}^x, u_{n_2 C}^y)$, and further denote the displacement of the entire strip as $\textbf{u}=(\textbf{u}_1,\textbf{u}_2,...,\textbf{u}_{N_2})$. To fully gap the band structure of the lattice, we introduce an on-site potential $\frac{1}{2}K'\textbf{u}\textbf{u}^T$ by embedding the lattice on a soft substrate.

Before calculating the elastic waves, we first derive the tension of a bond connecting two sites $i$ and $j$. The bond rest length is $l_0$ and it's unit vector is $\hat{n} = (\cos\theta,\sin\theta)$. We denote the relative displacement of the bond as $\Delta \vec u_{ij} = \vec u_j-\vec u_i$, and expand the tension in orders of it, with $\vec T_{i} = \vec F_{i}+\vec f_{i}+\mathcal{O}(\Delta u_{ij}^3)$. The leading order term is
\begin{eqnarray}\label{C0}
\vec F_{i} =  K(\hat{n}\hat{n})\Delta\vec u_{ij}
\end{eqnarray}
and the second order ones are
\begin{eqnarray}\label{C1.1}
 & {} & f_{i}^{x} =  \frac{K}{4l_0}\bigg[6\cos\theta\sin^2\theta (\Delta u_{ij}^{x})^2\nonumber \\
 & {} & -[3\sin(3\theta)-\sin\theta]\Delta u_{ij}^x\Delta u_{ij}^y-2\cos\theta(3\sin^2\theta-1)(\Delta u_{ij}^{y})^2\bigg]\nonumber \\
\end{eqnarray}
\begin{eqnarray}\label{C1}
 & {} & f_{i}^{y} = \frac{K}{4l_0}\bigg[6\sin\theta\cos^2\theta (\Delta u_{ij}^{y})^2\nonumber \\
 & {} & +[3\cos(3\theta)+\cos\theta]\Delta u_{ij}^x\Delta u_{ij}^y-2\sin\theta(3\cos^2\theta-1)(\Delta u_{ij}^{x})^2\bigg]\nonumber \\
\end{eqnarray}
Eqs.(\ref{C1.1}, \ref{C1}) are useful to derive the effective second harmonic feedback force.

We now apply an external driving at cell $g$ of the strip, denoted as ${F}_{n_1n_2}^{\rm ext}(t) = e^{i\omega t-ik_1n_1}{F}_{C}^y\delta_{n_2g}$. Thus, we further denote the external force as a $6N_2$-dimensional vector $\textbf{F}^{\rm ext}_g=(\textbf{F}^{{\rm ext}}_1, \textbf{F}^{{\rm ext}}_2, ..., \textbf{F}^{{\rm ext}}_{N_2})$. We assume $F_C^y$ is not too large, rendering all displacements $|\vec u_{A,B,C}|\ll l_{i=1,2,...,6}$, which further validates perturbation theory. Thus, we expand $\textbf{u}$ in orders of $F_C^y$, as $\textbf{u}=\textbf{u}^{(1)}+\textbf{u}^{(2)}+\mathcal{O}(F^3)$, where $\textbf{u}^{(1)}$ and $\textbf{u}^{(2)}$ are linear and 2nd harmonic modes. We define the frequency response function as 
\begin{eqnarray}\label{C3.1}
\textbf{G}(\omega,k_1) = [\textbf{D}(k_1)+(-m\omega^2+K'+i\eta\omega)\textbf{I}]^{-1}\qquad
\end{eqnarray}
where $\textbf{D}(k_1)=K\textbf{C}^\dag(k_1)\textbf{C}(k_1)$ is the dynamical matrix of the supercell strip. Thus, the linear mode is 
\begin{eqnarray}\label{C3}
\textbf{u}_{g}^{(1)}(\omega,k_1)=\textbf{G}(\omega,k_1)\textbf{F}^{\rm ext}(\omega,k_1).
\end{eqnarray}
The 2nd harmonic mode is 
\begin{eqnarray}\label{C4}
\textbf{u}_g^{(2)}(2\omega,2k_1)=\textbf{G}(2\omega, 2k_1)\textbf{f}^{(2)}(\textbf{u}_g^{(1)}),
\end{eqnarray}
where $\textbf{f}^{(2)}(\textbf{u}_g^{(1)})=(\textbf{f}_{1},\textbf{f}_{2},...,\textbf{f}_{N_2})$ is the second harmonic effective feedback force generated by the linear mode $\textbf{u}_g^{(1)}$, as shown in Eqs.(\ref{C1.1}, \ref{C1}).

The generic form of linear mode $\textbf{u}_g^{(1)}$ is given as follows:
\begin{eqnarray}\label{C5}
 & {} & \textbf{u}_{n_2,g}^{(1)} = \sum_{\alpha=1}^4a_{ g\alpha}\lambda_\alpha^{n_2}\bm{\phi}_\alpha \qquad 1\le n_2\le g\nonumber \\
 & {} & \textbf{u}_{n_2,g}^{(1)} = \sum_{\alpha=1}^4b_{ g\alpha}\lambda_\alpha^{n_2}\bm{\phi}_\alpha \qquad g+1\le n_2\le N_2,\qquad
\end{eqnarray}
where $\lambda_\alpha$, $\alpha=1,2,3,4$ are the eigenvalues of $[\textbf{D}(k_1,\lambda)+(-m\omega^2+K'+i\eta\omega)\textbf{I}]$, and $\bm{\phi}_\alpha$ are the corresponding $6\times 1$ eigenvectors. $\textbf{D}(k_1,\lambda)$ is the following $6\times 6$ matrix, 
\widetext
\begin{eqnarray}\label{C6}
\textbf{D}(k_1,\lambda) = \left(
\begin{array}{ccc}
\textbf{D}_1+\textbf{D}_3+\textbf{D}_4+\textbf{D}_5 & -\textbf{D}_1- e^{-ik_1} \textbf{D}_5 & -\textbf{D}_3-\lambda e^{-ik_1}\textbf{D}_4\\
-\textbf{D}_1-e^{ik_1}\textbf{D}_5 & \textbf{D}_1+\textbf{D}_2+\textbf{D}_5+\textbf{D}_6 & -\textbf{D}_2-\lambda \textbf{D}_6\\
-\textbf{D}_3-\lambda^{-1}e^{ik_1}\textbf{D}_4 & -\textbf{D}_2-\lambda^{-1}\textbf{D}_6 & \textbf{D}_2+\textbf{D}_3+\textbf{D}_4+\textbf{D}_6
\end{array}
\right).
\end{eqnarray}

\endwidetext
\noindent
$a_{g\alpha}$ and $b_{g\alpha}$, $\alpha = 1,2,3,4$ are constants determined by the open boundary condition at cell $1$,
\begin{eqnarray}\label{C7}
 & {} & (\textbf{D}_4e^{ik_1}, \textbf{D}_6, -\textbf{D}_4-\textbf{D}_6)\nonumber \\
 & {} & \cdot \left(
u^{(1)x}_{0,g A},
u^{(1)y}_{0,g A},
u^{(1)x}_{0,g B},
u^{(1)y}_{0,g B},
u^{(1)x}_{1,g C},
u^{(1)y}_{1,g C}
\right)^T=0,\qquad 
\end{eqnarray}
and the open boundary condition at cell $N_2$, 
\begin{eqnarray}\label{C8}
(\textbf{D}_3, \textbf{D}_2, -\textbf{D}_2-\textbf{D}_3)
\textbf{u}_{N_2,g}^{(1)}
=0.
\end{eqnarray}
Together with Eq.(\ref{C3}), we solve $a_{g\alpha}$ and $b_{g\alpha}$ to obtain the linear mode $\textbf{u}_g^{(1)}$. 



We then calculate the second harmonic mode based on Eq.(\ref{C4}). However, it is not easy to solve $\textbf{u}_g^{(2)}$, because the effective second harmonic feedback force, $\textbf{f}^{(2)}_{g}(2\omega, 2k_1)$ is applied at every cell. In order to simplify this problem, we consider the mode $\textbf{u}_{g'g}^{(2)}(2\omega,2k_1)$ in response to the second harmonic effective feedback force $\textbf{f}_{g'g}^{(2)}$ applied at a single cell $g'$, 
\begin{eqnarray}\label{C11}
\textbf{u}_{g'g}^{(2)}(2\omega,2k_1)=\textbf{G}(2\omega,2k_1)\textbf{f}^{(2)}_{g'g}(2\omega,2k_1).
\end{eqnarray}
The generic form of $\textbf{u}^{(2)}_{g'g}(2\omega, 2k_1)$ is given by
\begin{eqnarray}\label{C12}
 & {} & \textbf{u}^{(2)}_{n_2,g'g} = \sum_{\beta=1}^4c_{g'g\beta}\mu_{\beta}^{n_2}\bm{\varphi}_{\beta} \qquad 1\le n_2\le g'\nonumber \\
 & {} & \textbf{u}^{(2)}_{n_2,g'g} = \sum_{\beta=1}^4d_{g'g\beta}\mu_{\beta}^{n_2}\bm{\varphi}_{\beta} \qquad g'+1\le n_2\le N_2,\qquad
\end{eqnarray}
where $\mu_\beta$, $\beta = 1,2,3,4$ are the eigenvalues of $[\textbf{D}(2k_1,\mu)+(-4m\omega^2+K'+2i\eta\omega)\textbf{I}]$, and $\bm{\varphi}_\beta$ are the corresponding $6\times 1$ eigenvectors. $c_{g'g\beta}$ and $d_{g'g\beta}$ are constants determined by the boundary condition at cell $1$
\begin{eqnarray}\label{C14}
 & {} & (\textbf{D}_4e^{ik_1}, \textbf{D}_6, -\textbf{D}_4-\textbf{D}_6) \cdot\nonumber \\
 & {} & \left(
u^{(2)x}_{0,g'g A},
u^{(2)y}_{0,g'g A},
u^{(2)x}_{0,g'g B},
u^{(2)y}_{0,g'g B},
u^{(2)x}_{1,g'g C},
u^{(2)y}_{1,g'g C}
\right)^T=0,\nonumber \\
\end{eqnarray}
and open boundary condition at cell $N_2$
\begin{eqnarray}\label{C15}
(\textbf{D}_3, \textbf{D}_2, -\textbf{D}_2-\textbf{D}_3)\textbf{u}^{(2)}_{N,g'g}=0.
\end{eqnarray}
Together with Eq.(\ref{C11}), we solve $c_{g'g\beta}$ and $d_{g'g\beta}$ to obtain $\textbf{u}^{(2)}_{g'g}$. Finally, the second harmonic displacement is $\textbf{u}^{(2)}_{g}=\sum_{g'=1}^{N_2}\textbf{u}^{(2)}_{g'g}$.\\

\section{Numerical simulation of input local response function and output transmission susceptibility}
In the simulation, a finite kagome lattice which spans $40\vec a_1 \times (40-1)\vec a_2$ area in real space, is considered. The lattice is made of $40\times(40-1)$ unit cells with an additional layer of C-sites at the bottom. We connect the leftmost particles to the rightmost ones with harmonic springs to provide periodic boundary condition in the $\vec a_1$ direction, and we leave the C-sites of top and bottom boundaries free to realize open boundary conditions. By applying a vertical harmonic force $\vec F = (0,F)$ either on the top edge or bottom edge C-sites, we drive the lattice and we compute the displacement of mass points using a molecular dynamics scheme with damping. 
In order to make sure steady-state conditions are established, we wait $400\times (2\pi/\omega)$ before we make any displacement reading. We collect displacements $\vec u_{1C}(t)$ and $\vec u_{N_2 C}(t)$ on the two edges. By applying Fast Fourier Transformation (FFT), we convert  displacement time histories into their frequency spectra, $\vec u_{1 C}(\omega)$ and $\vec u_{N_2 C}(\omega)$. The elastic response is obtained via summation of multiple modes, 
\begin{eqnarray}
\vec u_{n_2 C}(\omega) = \vec u_{n_2 C}^{(1)}(\omega)+\vec u_{n_2 C}^{(2)}(2\omega)+...
\end{eqnarray}
where $\vec u_{n_2 C}^{(1)}(\omega) = (u_{n_2 C}^{x, (1)}(\omega), u_{n_2 C}^{y, (1)}(\omega))$ and $\vec u_{n_2 C}^{(2)}(2\omega) = (u_{n_2 C}^{x, (2)}(2\omega), u_{n_2 C}^{y, (2)}(2\omega))$. $u_{n_2 C}^{x, (1)}(\omega)$, $u_{n_2 C}^{y, (1)}(\omega)$, $u_{n_2 C}^{x, (2)}(2\omega)$ and $u_{n_2 C}^{y, (2)}(2\omega)$ are the amplitudes of $x$ and $y$ components of linear and second harmonic modes. The input linear response function is defined as $\chi_{\rm in+} = |u_{1 C}^{y, (1)}(\omega) |/ F$, and the output linear transmission susceptibility is defined as $\chi_{\rm out+}^{(1)} = |u_{N_2 C}^{y, (1)}(\omega) |/ F$. The output second harmonic transmission susceptibility is calculated through $\chi_{\rm out+}^{(2)} = |\vec u_{N_2 C}^{(2)}(2\omega) |/ F$.

%


\end{document}